\documentclass{CSML}
\pdfoutput=1

\usepackage{lastpage}
\newcommand{\dOi}{14(1:12)2018}
\lmcsheading{}{1--\pageref{LastPage}}{}{}%
{Dec.~01,~2016}{Jan.~30, 2018}{}

\usepackage{hyperref}
\hypersetup{hidelinks}

\usepackage{amssymb}
\usepackage{amssymb}
\usepackage{graphicx}
\usepackage{url}
\usepackage{graphicx}
 \usepackage{color}
 \usepackage{amsmath}
\usepackage{amsfonts}
\usepackage{amssymb}
\usepackage{url}
\usepackage{cite}
\usepackage{xspace}
\usepackage{graphicx}
\usepackage{subfigure}
\usepackage{color}
\usepackage{multicol}
\usepackage{fancyvrb}
\usepackage{todonotes}
\usepackage{tikz}
\usepackage{wrapfig}

\usetikzlibrary{positioning,mindmap,automata,fit,backgrounds,shapes, arrows}

\newcommand{\topolgy}{\mathcal{T}}

\newcommand{\LS}{\mathit{LS}}

\newcommand{\proc}{\pi}
\newcommand{\sys}{\mathcal{D}}

\newcommand{\flag}{\mathit{flag}}
\newcommand{\ind}{\mathit{Ind}}

\newcommand{\SYM}{\mathit{SYM}}
\newcommand{\SYN}{\mathit{SYN}}
\newcommand{\ASYN}{\mathit{ASYN}}
\newcommand{\CL}{\mathit{CL}}
\newcommand{\SC}{\mathit{SC}}

\newcommand{\va}{\mathit{val}}
\newcommand{\smt}{\mathit{SMT}}
\newcommand{\bool}{\mathit{Bool}}
\newcommand{\val}{\_val}

\newcommand{\statespace}{\mathrm{\Sigma}}
\newcommand{\states}{S}

\newcommand{\up}{up}

\newcommand{\tru}{\mathsf{true}}
\newcommand{\fals}{\mathsf{false}}
\newcommand{\mylambda}{\gamma}
\newcommand{\lambdai}{\gamma_i}

\newcommand{\lp}{\mathit{lp}}
\newcommand{\LP}{\mathit{LP}}
\newcommand{\token}{\mathit{tk}}
\newcommand{\inter}{\textsc{I}}

\newcommand{\col}{\mathit{color}}
\newcommand\remove[1]{\iffalse{#1}\fi}
\newcommand\change[1]{\color{black}{#1}}

\newcommand{\ltl}{\textsc{LTL}\xspace}
\newcommand{\ltlr}{\textsc{LTL}\xspace}

\newcommand\ltlF{\textsf{\textbf{F}}\,}
\newcommand\ltlG{\textsf{\textbf{G}}\,}
\newcommand\ltlU{\,\textsf{\textbf{U}}\,}
\newcommand\ltlX{\,\textsf{\textbf{X}}\,}
\newcommand{\Mod}[1]{\ \text{mod}\ #1}

\theoremstyle{plain}

\newcommand\removeCR[1]{}

\newtheorem{definition}[thm]{Definition}

\graphicspath{{figs/}}

\begin{document}

\title{Automated Synthesis of Distributed Self-Stabilizing 
Protocols}
\author[F. Faghih]{Fathiyeh~Faghih\rsuper{a}}
\address{\lsuper{a}College of Engineering, University of Tehran, Iran}
\email{f.faghih@ut.ac.ir }

\author[B. Bonakdarpour]{Borzoo~Bonakdarpour\rsuper{b}}
\address{\lsuper{b}Iowa State University, USA}
\email{borzoo@iastate.edu}

\author[S. Tixeuil]{S\'ebastien Tixeuil\rsuper{c}}
\address{\lsuper{c}LIP6, UPMC Sorbonne Universit\'{e}s, France}
\email{sebastien.tixeuil@lip6.fr}

\author[S. Kulkarni]{Sandeep~Kulkarni\rsuper{d}}
\address{\lsuper{d}Michigan State University, USA}
\email{sandeep@cse.msu.edu}


\titlecomment{A preliminary version of the paper has appeared in~\cite{fbtk16}}

\maketitle

\begin{abstract}

In this paper, we introduce an SMT-based method that automatically synthesizes 
a distributed {\em self-stabilizing} protocol from a given high-level 
specification and network topology. Unlike existing approaches, where synthesis 
algorithms require the {\em explicit} description of the set of legitimate 
states, our technique only needs the temporal behavior of the protocol. We 
extend our approach to synthesize {\em ideal-stabilizing} protocols, where every 
state is legitimate. We also extend our technique to synthesize {\em 
monotonic-stabilizing} protocols, where during recovery, each process can 
execute an most once one action. Our proposed methods are fully implemented and 
we report successful synthesis of well-known protocols such as Dijkstra's token 
ring, a self-stabilizing version of Raymond's mutual exclusion algorithm, 
ideal-stabilizing leader election and local mutual exclusion, as well as 
monotonic-stabilizing maximal independent set and distributed Grundy coloring.

\end{abstract}


\section{Introduction}
\label{sec:intro}

{\em Self-stabilization}~\cite{dij74} has emerged as one of the prime 
techniques for forward fault recovery. A self-stabilizing protocol satisfies 
two requirements: (1) {\em Convergence} ensures that starting from any 
arbitrary state, the system reaches a set of {\em legitimate states} (denoted 
in the sequel by $\LS$) with no external intervention within a finite number of 
execution steps, provided no new faults occur; (2) {\em Closure} indicates 
that the system remains in $\LS$ thereafter.

As Dijkstra mentions in his belated proof of self-stabilization~\cite{dij86}, 
designing self-stabilizing systems is a complex task. Proving the 
correctness of these algorithms is even more tedious. Thus, having access to 
automated methods (as opposed to manual techniques such as~\cite{da15}) for 
{\em synthesizing} correct self-stabilizing systems is highly desirable. 
However, synthesizing self-stabilizing protocols incurs high time and space 
complexity~\cite{ke13}. The techniques proposed in~\cite{bka12,ef11,ke14,ak09} 
attempt to cope with this complexity using heuristic algorithms, but none of 
these algorithms are complete; i.e., they may fail to find a solution although 
there exists one.

\subsection{Motivation}

Recently, Faghih and Bonakdarpour~\cite{fb15} proposed a sound and complete 
method to synthesize finite-state self-stabilizing systems based on SMT-solving. 
However, the shortcoming of this work as well as the techniques 
in~\cite{bka12,ef11,ke14} is that an {\em explicit} description of $\LS$ is 
needed as an input to the synthesis algorithm. The problem is that developing a 
formal predicate for legitimate states is not at all a straightforward task. 
For instance, the predicate for the set of legitimate states for Dijkstra's 
token ring algorithm with three-state machines~\cite{dij74} for three processes 
is the following:
\begin{align*}
\nonumber  \LS = (& (x_0+1 \equiv_3 x_1) \; \wedge \; (x_1+1 \not \equiv_3 
x_2)) \; \vee \\
\nonumber (& (x_1 = x_0) \; \wedge \; (x_1+1 \not \equiv_3 x_2 )) \; \vee 
\\
\nonumber (& (x_1+1 \equiv_3 x_0) \; \wedge \; (x_1+1 \not \equiv_3 
x_2)) \; \vee \\
\nonumber (& (x_0+1 \not \equiv_3 x_1) \; \wedge \; (x_1+1 \not \equiv_3 
x_0) \wedge (x_1+1 \equiv_3 x_2))
\end{align*}
where $\equiv_3$ denotes modulo 3 equality and variable $x_i$ belongs to 
process $i$. Obviously, developing such a predicate requires substantial 
expertise and insight and is, in fact, the key to the solution. Ideally, the 
designer should only express the basic requirements of the protocols (i.e., the 
existence of a unique token and its fair circulation), instead of an obscure 
predicate such as the one above.

\subsection{Contributions}

In this paper, we propose an automated approach to synthesize self-stabilizing 
systems given (1) the network topology, and (2) the high-level specification of 
legitimate states in the linear temporal logic (\ltl)~\cite{p77}. We also 
investigate automated synthesis of two important refinements of 
self-stabilization, namely \emph{ideal stabilization}~\cite{NT13} and 
\emph{monotonic stabilization}~\cite{YT10}. Ideally stabilizing protocols 
address two drawbacks of self-stabilizing protocols, namely exhibiting 
unpredictable behavior during recovery and poor compositional properties. In 
order to keep the specification as abstract as possible, the input \ltl 
formula may include a set of uninterpreted predicates. In designing 
ideal-stabilizing systems, the transition relation of the system and 
interpretation function of uninterpreted predicates must be found such that the 
specification is satisfied in every state.

Monotonic stabilization~\cite{YT10} relates to the behavior of a 
self-stabilizing system during stabilization, as it mandates a participating 
processor to change its output at most once after a transient fault occurs. So, 
a legitimate state is reached after at most one output change at every process. 
Intuitively, monotonic stabilization prevents unnecessary oscillations during 
stabilization, and guarantees recovery in a monotonic way (the system always 
moves closer to a legitimate state). Generic approaches to monotonic 
stabilization~\cite{YT10} require huge memory and time resources as the 
monotonic stabilization layer is added to an existing protocol. Finding specific 
monotonically stabilizing protocols that are memory and time efficient is 
notoriously difficult, yet highly appealing. These difficulties further 
motivate the need for developing methods that can automatically synthesize 
distributed self-, ideal-, and monotonic-stabilizing protocols.

\begin{figure}
	\vspace{-20pt}
	\begin{center}
		\scalebox{0.75}{
			\begin{tikzpicture}[auto, node distance=2cm]
			
			\tikzstyle{block} = [draw, rectangle];
			\tikzstyle{rblock}=[draw, shape=rectangle,rounded 
			corners=0.5em, fill=black, text=white, minimum height=3em, 
			minimum 
			width=5em];
			\node[block] (Topology) at (0, 3) {Network Topology};
			\node[block, align=center] (Type1) at (0, 5.2) 
			{Symmetry\\((a)symmetric)};
			\node[block, align=center] (Type2) at (3.5, 5.2) {Timing 
				Model\\((a)synchronous)};
			\node[block, align=center] (Type3) at (7.5, 5.2) 
			{Self-Stabilization\\(strong, weak, \\ ideal, monotonic)};
			\node[rblock, align=center] (SA) at (3.5, 3) {Synthesis};
			\node[block, align=center, below of = SA, node distance=2cm] 
			(LS){Legitimate Behavior\\(explicit/implicit)};
			\node[block, align=center, right of = SA, node distance=4cm] 
			(SS) 
			{Self-Stabilizing\\Protocol\\(Guarded Commands)};
			
			\draw[->,ultra thick, draw=blue] (Topology) to 
			(SA);
			\draw[->,ultra thick,draw=blue] (Type1) to (SA);
			\draw[->,ultra thick,draw=blue] (Type2) to (SA);
			\draw[->,ultra thick,draw=blue] (Type3) to (SA);
			\draw[->,ultra thick,draw=blue] (LS) to (SA);
			\draw[->,ultra thick,draw=blue] (SA) to (SS);
			
			\end{tikzpicture}
		}
	\end{center}
	\caption{Input and output of our synthesis method.}
	\label{fig:assess}
\end{figure}
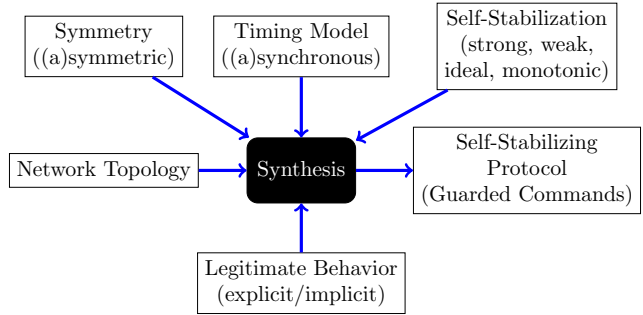

Our synthesis approach is inspired by bounded-synthesis~\cite{fs13}, where we 
transform the input specification into a set of SMT constraints. If the SMT 
instance is satisfiable, then a witness solution to its satisfiability encodes 
a distributed protocol that meets the input specification and topology. If the 
instance is not satisfiable, then we are guaranteed that no protocol that 
satisfies the input specification exists. The inputs and output of our 
synthesis method are depicted in Fig.~\ref{fig:assess}.

We also conduct several case studies using the model finder Alloy~\cite{j12}. 
In the case of self-stabilizing systems, we successfully synthesize 
Dijkstra's~\cite{dij74} (three-state machine) token ring and 
Raymond's~\cite{R89} mutual exclusion algorithms without explicit legitimate 
states as input. We also synthesize ideal-stabilizing leader election and local 
mutual exclusion (in a line topology) protocols, as well as 
monotonic-stabilizing distributed maximal independent set protocols and Grundy 
coloring.

\paragraph{\em Comparison to the conference version. } A preliminary version 
of this article appeared in the 36th International Conference  on Formal 
Techniques for Distributed Objects, Components, and Systems (FORTE'16). 
This article extends the conference version as follows:

\begin{itemize}
\item We extend our synthesis approach to synthesize monotonic-stabilizing 
protocols.

\item We conduct three case studies on synthesizing monotonic-stabilizing 
distributed maximal independent set, and Grundy coloring.

\end{itemize}
More precisely, 
Subsections~\ref{subsec:monotonic},~\ref{subsec:mon-syn}, 
and~\ref{sec:results-ms} are our added contributions.

\paragraph{\em Organization} In Sections~\ref{sec:model} and~\ref{sec:ss}, we 
present the preliminary concepts on the shared-memory model and  
self-stabilization. Section~\ref{sec:prob} formally states the synthesis 
problems. Formalization of timing models and symmetry in distributed 
programs are described in Section~\ref{sec:types}. In Section~\ref{sec:syn}, we 
describe our SMT-based technique, while Section~\ref{sec:results} is dedicated 
to our case studies. In Section~\ref{sec:disc}, we respond to some of the 
questions often raised about this line of work. We discuss the related work in 
Section~\ref{sec:related} and make concluding remarks 
Section~\ref{sec:conclusion}. 


\section{Model of Computation}
\label{sec:model}

\subsection{Distributed Programs}

Throughout the paper, let $V$ be a finite set of discrete {\em variables}. 
Each variable $v \in V$ has a finite domain $D_v$. A {\em state} is a 
mapping from each variable $v \in V$ to a value in its domain $D_v$. We call 
the set of all possible states the {\em state space}. A {\em transition} in the 
state space is an ordered pair $(s_0,s_1)$, where $s_0$ and $s_1$ are two 
states. We denote the value of a variable $v$ in state $s$ by $v(s)$.

\begin{definition}
 \label{def:process}
  A {\em process} $\proc$ over a set $V$ of variables is a tuple $\langle 
R_\proc,W_\proc,T_\proc \rangle$, where
\begin{itemize}
 \item $R_\proc \subseteq V$ is the {\em read-set} of $\proc$; i.e., variables 
that $\proc$ can read;

\item $W_\proc \subseteq R_\proc 
$ is the {\em write-set} of $\proc$; i.e., variables that $\proc$ can write, and

\item $T_\proc$ is the set of transitions of $\proc$, such that
$(s_0, s_1) \in T_\proc$ implies that for each variable $v \in V$, if $v(s_0) 
\neq v(s_1)$, then $v \in W_\proc$. \qed
\end{itemize}
\end{definition}
 
Notice that Definition~\ref{def:process} requires that a process can only 
change the value of a variable in its write-set (third condition), but not 
blindly (second condition). We say that a process $\proc = \langle 
R_\proc,W_\proc,T_\proc \rangle$ is {\em enabled} in state $s_0$, if there 
exists 
a state $s_1$, such that $(s_0, s_1) \in T_\proc$.

\begin{definition}
\label{def:prog}
A {\em distributed program} is a tuple $\sys = \langle \Pi_\sys, T_\sys\rangle$, 
where 
\begin{itemize}
 \item $\Pi_\sys$ is a set of processes over a common set $V$ of variables, such 
that:

\begin{itemize}
 \item for any two distinct processes $\proc_1, \proc_2 \in \Pi_\sys$, we 
have $W_{\proc_1} \cap W_{\proc_2} = \emptyset$;

\item for each process $\pi \in \Pi_\sys$ and each transition $(s_0, s_1) \in 
T_\pi$, the following {\em read restriction} holds:
\begin{align}
\nonumber \forall s'_0, s'_1: \Bigg(&\Big(\forall v \in R_\proc : \big(v(s_0) = 
v(s'_0) \; \wedge \; v(s_1) = v(s'_1)\big)\Big) \; \wedge \\ \label{eq:readdef}
& \Big(\forall v \not \in R_\proc : v(s'_0) = v(s'_1)\Big)\Bigg)  \implies 
(s'_0, s'_1) \in T_\proc
\end{align}
\end{itemize}
 
\item $T_\sys$ is the set of transitions and is the union of transitions 
  of all processes:
  \begin{equation}
    T_\sys = \bigcup_{\pi \in \Pi_\sys} T_\pi
    \tag*{\qEd}
  \end{equation}
\end{itemize} 
\end{definition}

\noindent Intuitively, the read restriction in Definition~\ref{def:prog} imposes 
the constraint that for each process $\proc$, each transition in $T_\proc$ 
depends only on reading the variables that $\proc$ can read. Thus, each 
transition forms an equivalence class in $T_\sys$, which we call a {\em group} 
of transitions. The key consequence of read restrictions is that during 
synthesis, if a transition is included (respectively, excluded) in $T_\sys$, 
then its entire corresponding group must be included (respectively, excluded) 
in $T_\sys$ as well. Also, notice that $T_\sys$ is defined in such a way that 
$\sys$ resembles an {\em asynchronous} distributed program, where process 
transitions execute in an {\em interleaving} fashion.

%
%
%
%

\paragraph{\em Example} 
Let $V = \{x_0, x_1, x_2\}$ be the set of variables, where $D_{x_0} = 
D_{x_1} =D_{x_2} =\{0,1,2\}$.  Let $\sys = \langle \Pi_\sys, T_\sys\rangle$ be 
the distributed program, where $\Pi_\sys = \{\proc_0, \proc_1, \proc_2 \}$. 
Each 
process $\proc_i$ ($0 \leq i \leq 2$) can write variable $x_i$. Also, 
$R_{\proc_0} = \{x_0,x_1\}$, $R_{\proc_1} = \{x_0,x_1, x_2\}$, and $R_{\proc_2} 
= \{x_1, x_2\}$. Notice that following Definition~\ref{def:prog} and read/write 
restrictions of 
$\proc_0$, the following (arbitrary) transitions 
\begin{align}
\nonumber t_1 = \big([&x_0 = 1,x_1=1,x_2 = 0], [x_0 = 2,x_1=1,x_2 = 0]\big)\\
\nonumber t_2 = \big([&x_0 = 1,x_1=1,x_2 = 2], [x_0 = 2,x_1=1,x_2 = 2]\big)
\end{align}
are in the same group, since $\proc_0$ cannot read $x_2$. This implies that if 
$t_1$ is included in the set of transitions of a distributed program, then so 
should be $t_2$. Otherwise, execution of $t_1$ by $\proc_0$ will depend on the 
value of $x_2$, which, of course, $\proc_0$ cannot read. 


\begin{definition}
 \label{def:computation}
 A {\em computation} of $\sys = \langle \Pi_\sys, T_\sys\rangle$ is an infinite 
sequence of states $\overline{s} = s_0s_1\cdots$, such that: (1)  for all $i 
\geq 0$, we have $(s_i, s_{i+1}) \in T_\sys$, and (2) if a computation reaches a 
state $s_i$, from where there is no state $\mathfrak{s} \neq s_i$, such that 
$(s_i,\mathfrak{s}) \in T_\sys$, then the computation stutters at $s_i$ 
indefinitely. Such a computation is called a {\em terminating computation}.\qed

 \end{definition}

\subsection{Predicates}
\label{sec:pred}
Let $\sys = \langle \Pi_\sys, T_\sys\rangle$ be a distributed program over a 
set $V$ of variables. The {\em global state space} of $\sys$ is the set 
of all possible global states of $\sys$: 
$$\statespace_\sys = \prod_{v \in V}D_v$$
The {\em local state space} of a process $\proc \in \Pi_\sys$ is the set of all 
possible local states of $\proc$, that is, the states that $\proc$ can fully 
read:
$$\statespace_\pi = \prod_{v \in R_\pi}D_v$$

\begin{definition}
An {\em interpreted global predicate} of a distributed program $\sys = \langle 
\Pi_\sys, T_\sys\rangle$ is a subset of $\statespace_\sys$ and an {\em 
interpreted local predicate} of a process $\proc \in \Pi_\sys$ is a subset of 
$\statespace_\proc$.\qed
\end{definition}

\begin{definition}
Let $\sys = \langle \Pi_\sys, T_\sys\rangle$ be a distributed program. An {\em 
uninterpreted global predicate} $\up$ is an uninterpreted Boolean function from 
$\statespace_\sys$. An {\em uninterpreted local predicate} $\lp$ is an 
uninterpreted Boolean function from $\statespace_\proc$, for some $\proc \in 
\Pi_\sys$.\qed
\end{definition}

The {\em interpretation} of an uninterpreted global predicate is a Boolean 
function from the set of all states: $$\up_{\inter}: \statespace_D 
\mapsto \{\tru, \fals\}$$ \noindent {\remove{$\up_{\inter}$ represents an 
interpreted global predicate that includes all states
mapped to true.}} Similarly, the interpretation of an uninterpreted local 
predicate for a process $\proc$ is a Boolean function: $$\lp_{\inter}: 
\statespace_\proc
\mapsto \{\tru, \fals\}$$ \noindent  {\remove{$\lp_{\inter}$ represents an 
interpreted local predicate that includes all local states
mapped to true.}} 
Throughout the paper, we use `uninterpreted predicate' to refer to either 
uninterpreted global or local predicate, and use `global (local) predicate' to 
refer to interpreted global (local) predicate.

\subsection{Topology}

A topology specifies the communication model of a distributed 
program.

\begin{definition}
\label{def:topology}

A {\em topology} is a tuple $\topolgy = \langle V,\lvert \Pi_\topolgy \rvert , 
R_\topolgy,W_\topolgy \rangle$, where

\begin{itemize}
\item $V$ is a finite set of finite-domain discrete variables;

\item $\lvert \Pi_\topolgy \rvert\ \in \mathbb{N}_{\geq 1}$ is the number of 
processes;

\item $R_\topolgy$ is a mapping $[0, \lvert \Pi_\topolgy \rvert-1] 
\mapsto 2^ V$ from a process index to its read-set, and

\item $W_\topolgy$ is a mapping $[0, \lvert \Pi_\topolgy \rvert-1] 
\mapsto 2^V$ from a process index to its write-set, such that $W_\topolgy(i) 
\subseteq R_\topolgy(i)$, for all $i \in [0, \lvert\Pi_\topolgy \rvert-1]	
$.\qed
\end{itemize} 
\end{definition}
{\remove{
 \paragraph{Example} \label{parag:topexample} The topology of our mutual exclusion problem is a tuple $\langle 
V,\lvert \Pi_\topolgy \rvert ,R_\topolgy,W_\topolgy \rangle$, where  
{ \begin{itemize}
    \item $V=\{c_0,c_1,c_2 \}$, with domains $D_{c_0} = D_{c_1} = D_{c_2} = \{0,1,2\}$,
    \item $\lvert \Pi_\topolgy \rvert\  = 3$,
    \item $R_\topolgy(0) = \{c_0,c_1\}$, 
$R_\topolgy(1) = \{c_0,c_1,c_2\}$,\\ $R_\topolgy(2) = \{c_1, 
c_2\}$, and
    \item $W_\topolgy(0) = \{c_0\}$, $W_\topolgy(1) = \{c_1\}$, and 
$W_\topolgy (2) = \{c_2\}$.
    \end{itemize}}
}}
\begin{definition}
A distributed program $\sys = \langle \Pi_\sys, T_\sys \rangle$ {\em has} 
topology $\topolgy = \langle V,\lvert 
\Pi_\topolgy \rvert ,R_\topolgy,W_\topolgy \rangle$ \; iff

\begin{itemize}
 \item each process $\proc \in \Pi_\sys$ is defined over $V$
 \item $|\Pi_\sys| = |\Pi_\topolgy|$
 \item there is a bijective function $g: [0, \lvert \Pi_\topolgy \rvert-1] 
\mapsto \Pi_\sys$ such that 
\begin{equation} \forall i \in [0, \lvert \Pi_\topolgy \rvert-1]: (R_\topolgy(i) = 
R_{g(i)}) \; \wedge \; (W_\topolgy(i) = W_{g(i)})
\tag*{\qEd}
\end{equation}
\end{itemize}
\end{definition}


\section{Formal Characterization of Self-, Monotonic-, and Ideal-Stabilization}
\label{sec:ss}

We specify the behavior of a distributed stabilizing program based on (1) 
the {\em functional} specification, and (2) the {\em recovery} specification. 
The functional specification is intended to describe what the program is 
required to do in a fault-free scenario (e.g., mutual exclusion or leader 
election). The recovery behavior stipulates Dijkstra's idea of 
self-stabilization in spite of distributed control~\cite{dij74}.

\subsection{The Functional Behavior}

We use \ltl~\cite{p77} to specify the functional behavior of a stabilizing 
program. Since LTL is a commonly-known language, we refrain from presenting its 
syntax and semantics ($\ltlF$, 
$\ltlG$, $\ltlX$, and $\ltlU$ denote the `finally', `globally', `next', 
and `until' operators, respectively). In our framework, an \ltl\ formula 
may include uninterpreted predicates. Thus, we say that a program $\sys$ 
satisfies an \ltl\ formula $\varphi$ from an initial state in the set $I$, 
and write $\sys,I \models \varphi$ \; iff \; there exists an 
interpretation function for each uninterpreted predicate in 
$\varphi$, such that all computations of $\sys$, starting from a state in 
$I$ satisfy $\varphi$. Also, the semantics of the satisfaction relation is 
the standard semantics of \ltl\ over Kripke structures (i,e., computations of 
$\sys$ that start from a state in $I$).

\paragraph{\em Example~\ref{subsec:ex1}} \label{subsec:ex1} Consider the 
problem of {\em token passing} in a ring topology (i.e., token ring), where 
each process $\proc_i$ has a variable $x_i$ with the domain 
$D_{x_i}=\{0,1,2\}$. This problem has two functional requirements:

\begin{description}
 \item[Safety]  The {\em safety} requirement for this problem is that in each 
state, only one process has the token and, hence, can execute. To formulate 
this requirement, we assume that each process $\proc_i$ is associated with a 
local uninterpreted predicate $\token_i$, which shows whether $\proc_i$ is 
enabled. Let $\LP = \{\token_i \; \mid \; 0 \le i < n \}$. Thus, a process 
$\proc_i$ can execute a transition, if and only if $\token_i$ is true.  The 
\ltl formula, $\varphi_{\mathbf{TR}}$, expresses the above requirement for 
a ring of size $n$:
$$\varphi_{\mathbf{TR}} = \forall i \in [0, n-1]: \token_i \iff \big(\forall 
\va \in \{0,1,2\}: (x_i=\va) \Rightarrow \ltlX (x_i \neq \va)\big)$$

Using the set of uninterpreted predicates, the safety requirement can be 
expressed by the following \ltl formula:
$$\psi_{\mathbf{safety}} = \exists i \in [0, n-1]: (\token_i \; \wedge 
\; \forall j \neq i: \neg \token_j)$$

\noindent Note that although safety requirements generally need the 
$\ltlG$ operator, we do not need it, as every state in a stabilizing 
system can be an initial state.

\item[Fairness] This requirement express that for every process $\proc_i$ and 
starting from each state, the computation should reach a state, where 
$\proc_i$ is enabled:
$$\psi_{\mathbf{fairness}} = \forall i \in [0, n-1]: (\ltlF \token_i)$$ 
Another way to guarantee this requirement is that 
processes get enabled in a clockwise order in the ring, which can be formulated 
as follows:
$$\psi_{\mathbf{fairness}} = \forall i \in [0, n-1]:  
(\token_i \Rightarrow \ltlX \token_{(i+1 \Mod n)})$$

\noindent Note that the latter approach is a stronger constraint, and 
would prevent us to synthesize bidirectional protocols, such as Dijkstra's 
three-state solution.

\end{description}
Thus, the functional requirements of the token ring protocol is
$$\psi_{\mathbf{TR}}=\psi_{\mathbf{safety}} \; \wedge \; 
\psi_{\mathbf{fairness}}$$
Observe that following Definition~\ref{def:computation}, $\psi_{\mathbf{TR}}$ 
ensures deadlock-freedom as well.

\paragraph{\em Example 3.2} Consider the problem of {\em local mutual 
exclusion} on a line topology, where each process $\proc_i$ has a Boolean 
variable $x_i$. The requirements of this problem are as follows:

\begin{description}

\item[Safety] In each state, (1) at least one process is enabled, that is, 
(i.e., deadlock-freedom), and (2) no two neighboring processes are enabled 
(i.e., local mutual exclusion). To formulate these requirements, we associate 
with each process $\proc_i$ a local uninterpreted predicate $\token_i$, which 
is true when $\proc_i$ is enabled:
$$\varphi_{\mathbf{LME}} = \forall i \in [0, n-1]: \token_i \iff \Big((x_i 
\Rightarrow \ltlX \neg x_i) \; \wedge \; (\neg x_i \Rightarrow \ltlX 
x_i)\Big)$$
Thus, $\LP=\{\token_i \; \mid \; 0 \le i < n \}$ and the safety requirement can be 
formulated by the following \ltl formula:
$$\psi_{\mathbf{safety}} = (\exists i \in [0, n-1]: \token_i) \; \wedge 
\; (\forall i \in [0, n-2]: \neg (\token_i \wedge \token_{(i+1)}))$$

\item[Fairness]
Each process {\remove{$\proc_i$}} is eventually enabled:
$$\psi_{\mathbf{fairness}} = \forall i \in [0, n-1]: (\ltlF \token_i)$$

\end{description}
Thus, the functional requirement of the local mutual exclusion protocol is   
$$\psi_{\mathbf{LME}}=\psi_{\mathbf{safety}} \; \wedge 
\; \psi_{\mathbf{fairness}}$$

\subsection{Self-Stabilization}

A {\em self-stabilizing system}~\cite{dij74} is one that always recovers a good 
behavior (typically, expressed in terms of a set of {\em legitimate states}), 
starting from any arbitrary initial state.

\begin{definition}
\label{def:ss}
A distributed program $\sys = \langle \Pi_\sys, T_\sys\rangle$ is {\em 
self-stabilizing} for \ltl functional specification $\psi$ \; iff \; there 
exists a global predicate $\LS$ \footnote{$\LS$ is a set of states, which is 
implicitly specified in terms of an LTL formula. When it appears on 
the left-hand side of $\models$, it means that the formula on the right-hand 
side needs to hold in $\LS$, rather than in the whole state-space $\Sigma_D$. 
When $\LS$ appears in right-hand side of $\models$, it is part of the formula, 
which we need to hold.} (called the set of {\em legitimate states}), such that: 

\begin{itemize}
\item {\em Functional behavior:} $\sys, \LS \models \psi$

\item {\em Strong convergence:} $\sys,\statespace_\sys \models \ltlF \LS$

\item {\em Closure:} $\sys,\statespace_\sys \models (\LS \Rightarrow \ltlX \LS)$
\qed
\end{itemize}
\end{definition}

Notice that the strong convergence property ensures that starting from any 
state, any computation converges to a legitimate state of $\sys$ within a 
finite number of steps. The closure property ensures that execution of the 
program is closed in the set of legitimate states. In the sequel, we will omit 
the state space $\statespace_\sys$ and \ltl specification $\psi$, when they are 
clear from the context or they are irrelevant. 

There exist several results on impossibility of distributed 
self-stabilization (e.g., in token circulation and leader 
election in anonymous networks~\cite{h90}). Thus, weaker forms of 
stabilization have been introduced in the literature of distributed computing. 
One example is {\em weak-stabilizing} distributed programs~\cite{g01}, which is defined as follows.

\begin{definition}
\label{def:wss}

A distributed program $\sys = \langle \Pi_\sys, T_\sys\rangle$ is {\em 
weak-stabilizing} for a set $\LS$ of {\em legitimate states} \; if and 
only if \; the following conditions hold:

\begin{itemize}
\item {\em Weak convergence:} For each state $s_0$ in the state space of $\sys$, 
there exists a computation $\overline{s} = s_0s_1\cdots$ of $\sys$, where there 
exists $i\geq 0$, such that $s_i \in \LS$.\footnote{Observe that weak 
convergence cannot be expressed in LTL.}

\item {\em Functional behavior:} As defined in Definition~\ref{def:ss}

\item {\em Closure:} As defined in Definition~\ref{def:ss}.\qed
\end{itemize}
\end{definition}

Notice that unlike strong self-stabilizing programs, in a weak-stabilizing 
program, there may exist execution cycles outside the set of legitimate states. 
In the rest of the paper, we use `strong self-stabilization' (respectively, 
`strong convergence') and `self-stabilization' (respectively, `convergence') 
interchangeably.

\subsection{Ideal-Stabilization}

Self-stabilization does not predict program behavior during recovery, which may 
be undesirable for some applications. A simple way to integrate program 
behavior during recovery is to include it in the specification itself. This 
way, the protocol must ensure that every configuration in the specification is 
legitimate (so, the only recovery behaviors are those included in the 
specification). Such a protocol is called \emph{ideal stabilizing}~\cite{NT13}.

\begin{definition}
\label{def:iss}
Let $\psi$ be an \ltlr specification and $\sys = \langle \Pi_\sys, 
T_\sys\rangle$ be a distributed program. We say that $\sys$ is {\em ideal 
stabilizing} for $\psi$ \; iff \;  $\sys,\statespace_\sys \models \psi$.
\qed
\end{definition}

The existence of ideal stabilizing protocols for \ltlr specifications (that only 
mandate legitimate states) is an intriguing question, as one has to find a 
``clever'' set of transitions and an interpretation function for every 
uninterpreted predicate (if included in the specification), such that the system 
satisfies the specification at all times. Note that there is a specification 
for every system to which it ideally stabilizes~\cite{NT13}, and that is the 
specification that includes all of the system computations ($\psi=\tru$). In 
this paper, we do the reverse; meaning that getting an \ltlr specification 
$\psi$, we synthesize a distributed system that ideally stabilizes to~$\psi$.

\subsection{Monotonic-Stabilization}
\label{subsec:monotonic}

Monotonic stabilization~\cite{YT10} also relates to prescribing program behavior 
during recovery, as it requires every process to change its variable at most once 
after a transient fault occurs. This simple requirement induces desirable 
properties for fault recovery. For example, processes cannot go back and 
forth between states: once a variable  has been changed, it remains so until a 
legitimate state is reached, improving stability while recovering.

A generic approach to monotonic stabilization~\cite{YT10} is for each process to 
collect variable  information at some distance that depends on the considered 
problem, and change its variable  only if it is absolutely sure that it should do 
so. The space and time required to implement such a scheme is huge, even for 
relatively simple problems. In order to design a viable monotonically 
stabilizing protocol, a problem specific approach is necessary. This indeed 
makes manual design of monotonic-stabilizing protocols a tedious task.

\begin{definition}
A distributed program $\sys = \langle \Pi_\sys, T_\sys\rangle$ is {\em 
monotonic-stabilizing} \; iff 

\begin{itemize}
 \item $\sys$ is self-stabilizing with some set $\LS$ of legitimate states, and
 
 \item for every (recovery) computation $\overline{s} = s_0s_1\cdots s_n$ of 
$\sys$, where for all $i \in [0, n-1]$, we have $s_i \in \neg \LS$ and $s_n \in 
\LS$, the following holds:
\begin{align*}
\exists j \in [0, n-1]: \exists \proc \in \Pi_\sys: \exists v \in 
W_\proc: v(s_j) \neq v(s_{j+1}) \; \Rightarrow \\
\forall k \in [0, n-1]-\{j\}:\forall v \in W_\proc: v(s_k) = v(s_{k+1})
  \tag*{\qEd}
\end{align*}
\end{itemize}
\end{definition}






\section{Problem Statement}
\label{sec:prob}

Our goal is to develop a synthesis algorithm that takes as input (1) system 
topology, and (2) two \ltlr formulas $\varphi$ and $\psi$ that involve a set 
$\LP$ of uninterpreted predicates, and generate as output a self-, monotonic-,  
or ideal-stabilizing protocol. For instance, in token passing on a ring, 
$\psi_{\mathbf{TR}}$ includes safety and fairness, which should hold in the set 
of legitimate states, while $\varphi_{\mathbf{TR}}$ is a general requirement 
that we specify on every uninterpreted predicate $\token_i$. Since in the case 
of self-stabilizing systems, we do not get $\LS$ as a set of states (global 
predicate), we refer to our problem as ``synthesis of self-stabilizing systems 
with {\em implicit} $\LS$''.

\begin{center}
\fbox{\rule{1mm}{0mm}
\begin{minipage}[t]{.95\columnwidth}
\label{prb:smt2}

{\bf Problem statement 1 (self/monotonic-stabilization). } Given is 
\begin{enumerate}
\item a topology 
$\topolgy = \langle V,\lvert \Pi_\topolgy \rvert , R_\topolgy, W_\topolgy 
\rangle$;
\item two \ltlr formulas $\varphi$ and $\psi$ that involve a set $\LP$ of  
uninterpreted predicates.
\end{enumerate}
The synthesis algorithm is required to identify as output (1) a distributed 
program $\sys = \langle \Pi_\sys, T_\sys \rangle$, (2) an interpretation 
function for every local predicate $\lp \in \LP$, and (3) the global state 
predicate $\LS$, such that $\sys$ has 
topology $\topolgy$,  $\sys,\statespace_\sys \models \varphi$, and $\sys$ is 
self/monotonic-stabilizing for $\psi$.
\end{minipage}
}
\end{center}

\begin{center}
\fbox{\rule{1mm}{0mm}
\begin{minipage}[t]{.95\columnwidth}
\label{prb:smt3}
{\bf Problem statement 2 (ideal-stabilization). } Given is 

\begin{enumerate}
\item a topology 
$\topolgy = \langle V,\lvert \Pi_\topolgy \rvert , R_\topolgy, W_\topolgy 
\rangle$
\item two \ltlr formulas $\varphi$ and $\psi$ that involve a set $\LP$ of  
uninterpreted predicates.
\end{enumerate}
  The synthesis algorithm is required to generate as output (1) a distributed program 
  $\sys = \langle \Pi_\sys, T_\sys \rangle$, and (2) an interpretation function for every local predicate $\lp \in \LP$, such that $\sys$ has 
  topology $\topolgy$ and  $\sys,\statespace_\sys \models (\varphi \wedge \psi)$.
\end{minipage}
}
\end{center}


\section{Timing Models and Symmetry in Distributed Programs}
\label{sec:types}

We would like our synthesis solution to also take as input the timing model as 
well as symmetry requirements among processes. These constraints are defined 
in Subsections~\ref{subsec:timing} and~\ref{subsec:symmetry}.

\subsection{Timing Models}
\label{subsec:timing}

Two commonly-considered timing models in the literature of distributed 
computing are {\em synchronous} and {\em asynchronous} programs~\cite{l96}. In 
an asynchronous distributed program, every transition of the program is a 
transition of one and only one of its processes (central daemon model). 

\begin{definition}
 A distributed program $\sys = \langle \Pi_\sys, T_\sys \rangle$ is {\em 
asynchronous} \; if and only if \; the following condition holds:
\begin{align}
\label{eq:async}
\ASYN \; = \; & \forall (s_0, s_1) \in T_\sys: \Big((\exists \proc \in 
\Pi_\sys: (s_0, s_1) \in T_\proc) \; \vee  \nonumber \\
& ((s_0 = s_1) \; \wedge \; \forall \proc 
\in \Pi_\sys: \forall \mathfrak{s}: (s_0, \mathfrak{s}) \not \in T_\proc)\Big)
\end{align}
\end{definition}
Thus, the set of transitions of an asynchronous program is simply the union of 
transitions of all its processes. That is, $$T_\sys = \bigcup_{\pi \in 
\Pi_\sys} T_\pi$$ An asynchronous distributed program resembles 
a system, where process transitions execute in an {\em interleaving} fashion.

In a synchronous distributed program, on the other hand, in every step, all 
enabled processes have to take a step simultaneously.

\begin{definition}
A distributed program $\sys = \langle \Pi_\sys, T_\sys \rangle$ is {\em 
synchronous} \; if and only if \; the following condition holds:
\begin{align}
\label{eq:sync}
\nonumber \SYN \, = \, & \forall (s_0, s_1) \in T_\sys: \,  \forall 
\proc \in \Pi_\sys: \\ & \big(\exists \mathfrak{s}: ((s_0, \mathfrak{s}) \in 
T_\proc) \, \wedge \, \forall v \in W_\proc:v(s_1) = v(\mathfrak{s})\big) \; 
\vee \\
& \big(\forall \mathfrak{s}: ((s_0, \mathfrak{s}) \not \in T_\proc) \,
\wedge \, \forall v \in W_\proc:v(s_0) = v(s_1)\big) \nonumber 
          \tag*{\qEd}
\end{align}
\end{definition}
In other words, a distributed program is synchronous, if and only if each 
transition $(s_0, s_1) \in T_\sys$ is obtained by execution of all enabled 
processes (the ones that have a transition starting from $s_0$). Hence, 
the value of the variables in their write-sets change in $s_1$ accordingly. 
Also, for all non-enabled processes, the value of the variables in their 
write-sets do not change from $s_0$ to $s_1$.

\subsection{Symmetry}
\label{subsec:symmetry}

Symmetry in distributed programs refers to similarity of behavior of 
different processes. 

\begin{definition}
\label{def:sym}
A distributed program $\sys = \langle \Pi_\sys, T_\sys\rangle$ is called {\em 
symmetric} \; if and only if \; for any two distinct processes $\proc, \proc' 
\in 
\Pi_\sys$, there exists a bijection $f: R_\proc \rightarrow 
R_{\proc'}$, such that the following condition holds:
\begin{align}
\label{eq:sym}
  \begin{array}{r@{}l}
\SYM \, = \ & \forall (s_0, s_1) \in T_\proc: \;  \exists (s'_0, s'_1) 
\in T_{\proc'}: \\
& \big(\forall v \in R_\proc: (v(s_0) = f(v)(s'_0))\big) \; \wedge  
\big(\forall v \in 
W_\proc: (v(s_1) = f(v)(s'_1))\big)
  \end{array}
\end{align}
\qed
\end{definition}
In other words, in a symmetric distributed program, the transitions of a 
process can be determined by a simple variable mapping from another process. A 
distributed program is called {\em asymmetric} if it is not symmetric.


\section{SMT-based Synthesis Solution}
\label{sec:syn}

Our technique is inspired by our SMT-based method in~\cite{fb15}. In 
particular, we transform the problem input into an {\em SMT instance}. An SMT 
instance consists of two parts: (1) a set of {\em entity} declarations (in 
terms of sets, relations, and functions), and (2) first-order modulo-theory 
{\em constraints} on the entities. An SMT-solver takes as input an SMT instance 
and determines whether or not the instance is satisfiable. If so, then a 
witness generated by the SMT solver is the answer to our synthesis problem. We 
describe the SMT entities obtained in our transformation in 
Subsection~\ref{subsec:entities}. SMT constraints appear 
in Subsections~\ref{subsec:generalcon}-~\ref{subsec:specific-SMT}. Note that 
using our approach in~\cite{fb15}, we can synthesize different systems 
considering types of timing models (i.e., synchronous and asynchronous), 
symmetric and asymmetric, as well as strong- and weak-stabilizing protocols. 

\subsection{SMT Entities}
\label{subsec:entities}

Recall that the inputs to our problems include a topology $\topolgy = \langle 
V,\lvert \Pi_\topolgy \rvert , R_\topolgy, W_\topolgy \rangle$, and two \ltlr formulas on a set  $\LP$ of 
uninterpreted predicates. Let $D = \langle \Pi_\sys, 
T_\sys\rangle$ denote a distributed program that is a solution to our problem. In our SMT instance, we 
include:
\begin{itemize}

\item A set $D_v$ for each $v \in V$, which contains the elements in the 
domain of $v$.

\item A set $\bool$ that contains the elements $\tru$ and $\fals$.

\item A set called $\states$, whose cardinality is $\biggr\rvert\prod\limits_{v 
\in V}D_v\biggr\rvert$. This set represents the state space of the synthesized 
distributed program.

\item An uninterpreted function $v\val$ for each variable $v$; i.e., $v\val: 
\states \mapsto D_v$.

\item An uninterpreted function $lp\val$ for each uninterpreted predicate $\lp 
\in \LP$; i.e, $\lp\val: \states \mapsto \bool$.

\item  An uninterpreted relation $T_i \subseteq \states \times \states$ that represents the transition relation for process 
$\proc_i$ in the synthesized program.

\item An uninterpreted function $\mylambda$, from each state to a natural 
number ($\mylambda \; : \; \states \mapsto \mathbb{N}$). This function is used to capture convergence to the set of legitimate states. 

\item An uninterpreted function $\LS \, : \, \states \mapsto \bool$.

\end{itemize}

\noindent The last two entities are only included in the case of Problem 
Statement 1.

\paragraph{\em Example} For Example~\ref{subsec:ex1}, we include the following 
SMT entities:

\begin{itemize}

\item $D_{x_0} = D_{x_1} = D_{x_2} = \{0,1,2\}$, $\bool = \{\tru,\fals\}$, set $\states$, where $\lvert \states \rvert = 27$ 

\item $x_0\val: \states \mapsto D_{x_0}$, $x_1\val : \states 
\mapsto D_{x_1}$, $x_2\val: \states \mapsto D_{x_2}$

\item $T_0 \subseteq \states \times \states$, $T_1 \subseteq \states \times \states$, $T_2 \subseteq \states \times \states$
, $\mylambda: \; \states \mapsto \mathbb{N}$ 
, $\LS \, : \, \states \mapsto \bool$
\end{itemize}

\subsection{General SMT Constraints}
\label{subsec:generalcon}

\subsubsection{State Distinction}
\label{subsec:statedistinction}
 
Any two states differ in the value of some variable:
 \begin{align}  \label{eq:statedist}
 \forall s,s' \in \states \; : \; & (s \neq s') \; \Rightarrow \;  
(\exists v \in V \; : \; v\val(s) \neq v\val(s'))
 \end{align}

\subsubsection{Local Predicates Constraints}
             
Let $\LP$ be the set of uninterpreted predicates used in formulas $\varphi$ and 
$\psi$. For each uninterpreted local predicate $\lp_{\proc}$, we need to ensure 
that its interpretation function is a function of the variables in the read-set 
of $\proc$. To guarantee this requirement, for each $\lp_{\proc} \in \LP$, we 
add the following constraint to the SMT instance:
\begin{align}
\nonumber \forall s,s' \in \states : ~ & \big(\forall v \in R_\proc : (v(s) = 
v(s'))  \; \Rightarrow \; (\lp_\proc(s) = \lp_\proc(s'))\big) 
\end{align}
            
\paragraph{\em Example} For Example~\ref{subsec:ex1}, we add the following 
constraint for process $\proc_1$: 
\begin{align}
  \begin{array}{r@{}l}
\forall s,s' \in \states: \big((x_0(s) = x_0(s')) \, \wedge \, 
(x_1(s) = x_1(s')) \, \wedge \, (x_2(s) = x_2(s'))\big)  \; \Rightarrow \\ 
(\token_1(s) = \token_1(s')) 
  \end{array}
\end{align}
     
\subsubsection{Constraints for an Asynchronous System}
\label{subsubsec:async}
      
To synthesize an asynchronous distributed program, we add the following 
constraint for each transition relation $T_i$:

\begin{align} \label{con:async}
\forall (s,s') \in T_i \; : \; &  \forall v \notin W_\topolgy(i) \; : \; 
v\val(s) = v\val(s') 
\end{align}
Constraint~\ref{con:async} ensures that in each relation $T_i$, only process 
$\proc_i$ can take a transition. By introducing $\lvert \Pi_\topolgy \rvert\ $ 
transition relations, we consider all possible interleaving of processes 
taking transitions. Note that this constraint is a formulation of the third item 
in Definition~\ref{def:process}.
     
\subsubsection{Read Restrictions} 
  
To ensure that $\sys$ meets the read restrictions given by $\topolgy$ and  
Definition~\ref{def:prog}, we add the following constraint for each process 
index:
\begin{align}
  \begin{array}{r@{}l}
\forall (s_0, s_1) \in T_i : \; \forall s'_0, s'_1: ~ \Big( 
&\big(\forall v \in R_\proc : (v(s_0) = v(s'_0) \; \wedge \; v(s_1) = 
v(s'_1))\big) \; \wedge \\
&\big(\forall v \not \in R_\proc : v(s'_0) = v(s'_1)\big)\Big)  \Rightarrow 
(s'_0, s'_1) \in T_i
  \end{array}
 \label{eq:readcon}
\end{align}
            
\subsection{Specific SMT Constraints for Self- and Ideal-Stabilizing Problems}
\label{subsec:specific-SMT}

Before presenting the constraints specific to each of our problem statements, we 
 present the formulation of an \ltlr formula as an SMT constraint. We use this 
formulation to encode the $\psi$ and $\varphi$ formulas (given as input) as 
$\psi_\smt$ and $\varphi_\smt$, and add them to the SMT instance. {\remove{(as 
discussed in Sections~\ref{subsubsec:impLS} and~\ref{subsec:ideal-syn}).}}

 \subsubsection{SMT Formulation of an \ltlr Formula}
 \label{subsec:ltlrform}

SMT formulation of an \ltl\ formula is presented in~\cite{fs13}. Below, we 
briefly discuss the formulation of \ltl formulas without nested temporal 
operators. For formulas with nested operators, the formulation based on 
universal co-B\"{u}chi automata~\cite{fs13} needs to be applied.

 \paragraph{SMT formulation of $\ltlX$: {\remove{Operator:}}}
 \label{subsec:SMTX}  
A formula of the form $\textit{\ltlX} P$ is translated to an SMT constraint as below~\footnote{\change{Note that for a formula $P$, $P(s)$ is acquired  by replacing each variable $v$ with $v(s)$.}}:
\begin{align}
\forall s,s' \in \states \; : \; \forall i \in [0, |\Pi_\topolgy|-1] \; : \; 
(s,s') \in T_i \; \Rightarrow \; P(s')
\end{align}

 \paragraph{SMT formulation of $\ltlU$: {\remove{Operator:}}}
 \label{subsec:SMTF}
 Inspired by {\em bounded synthesis}~\cite{fs13}, for each formula of the 
form $P \ltlU Q$, we define an uninterpreted function $\lambdai: 
\states \mapsto \mathbb{N}$ and add the following constraints to the 
SMT instance:
 \begin{align} 
  \label{eq:F}
\forall s,s' \in \states \; : \; &\forall i \in [0, |\Pi_\topolgy|-1]: 
\neg Q (s) \; \wedge \; (s,s') 
 \in T_i \; \Rightarrow (P(s) \; \wedge \; \lambdai(s') > \lambdai(s))  
\end{align}
 \begin{align}
 \label{eq:Fnodeadlock} 
  \forall s \in \states \; : \; \neg Q (s) \; \Rightarrow \; \exists  i \in [0, 
 |\Pi_\topolgy|-1] \; : \; \exists s' \in \states \; 
 : \; (s,s') \in T_i
  \end{align}
 
The intuition behind Constraints~\ref{eq:F} and~\ref{eq:Fnodeadlock} can be 
understood easily. If we can assign a natural number to each state, such 
that along each outgoing transition from a state in $\neg Q$, the number 
is strictly increasing, then the path from each state in $\neg Q$ should 
finally reach $Q$ or get stuck in a state, since the size of state space 
is finite. Also, there cannot be any loops whose states are all in $\neg 
Q$, as imposed by the annotation function.
 Finally, Constraint~\ref{eq:Fnodeadlock} ensures that there is no deadlock 
state in $\neg Q$ states.


\subsubsection{Synthesis of Self-Stabilizing Systems}
\label{subsec:imp-syn}

In this section, we present the constraints specific to synthesizing self-stabilizing systems.

\paragraph{Closure ($\CL$):}

The formulation of the closure constraint in our SMT instance is as follows:
\begin{align} \label{eq:closure}
\forall s,s' \in \states \; : \; \forall i \in [0, |\Pi_\topolgy|-1] \; : \; 
(\LS (s) \, \wedge \, (s,s') \in T_i) \; \Rightarrow \; \LS(s')
\end{align}

\paragraph{Strong Convergence ($\SC$):}
\label{subsubsec:strongconv}

Similar to the constraints presented in Section~\ref{subsec:SMTF}, our SMT 
formulation for $\SC$ is a simplification of Constraints~\ref{eq:F} 
and~\ref{eq:Fnodeadlock} (recall that $\ltlF \LS = \tru \ltlU \LS$):
\begin{align} 
\label{eq:sc2}
& \forall s,s' \in \states: \forall i \in \{0 \cdots|\Pi_\topolgy|-1\}: \neg 
\LS (s) \, \wedge \, (s,s') 
\in T_i \; \Rightarrow \; \mylambda(s') > \mylambda(s)  \\
& \label{eq:nodeadlock} 
 \forall s \in \states \; : \; \neg \LS (s) \; \Rightarrow \; \exists  i \in 
\{0 \cdots |\Pi_\topolgy|-1\} \; : \; \exists s' \in \states \; 
: \; (s,s') \in T_i
 \end{align}

\paragraph{General Constraints on Uninterpreted Predicates:}
\label{subsubsec:gen}

 As mentioned in Section~\ref{sec:prob}, one of the inputs to our problem is an 
\ltlr formula, $\varphi$ describing the role of uninterpreted predicates. 
Considering $\varphi_{\smt}$ to be the SMT formulation of  $\varphi$, we add the following SMT constraint to the SMT instance:
  \begin{align}
  \label{eq:general}
  \forall s \in \states \; : \; \varphi_{\smt}
  \end{align}

 \paragraph{Constraints on $LS$:}
 \label{subsubsec:impLS} 
 
 Another input to our problem is the \ltlr formula $\psi$ that includes 
requirements, which should hold in the set of legitimate states. We formulate 
this formula as SMT constraints using the method discussed in 
Section~\ref{subsec:SMTF}. Considering $\psi_{\smt}$ to be the SMT formulation 
of the $\psi$ formula, we add the following SMT constraint to the SMT instance:
\begin{align}
\label{eq:impLS}
\forall s \in \states \; : \; \LS(s) \; \Rightarrow \; \psi_{\smt}
\end{align}

\paragraph{\em Example} Continuing with Example~\ref{subsec:ex1}, we add the 
following constraints to encode $\varphi_{\mathbf{TR}}$:
\begin{align}
\nonumber
\forall s \in \states \; : \; \forall i \in [0, n-1] \; : \; \token_i(s) 
\iff & (\forall j \in [0, n-1] \; : \; j \neq i \; \Rightarrow \; \\ 
\nonumber & \nexists s' \in \states \; : \;  (s,s') \in T_j )
\end{align}
  \noindent Note that the asynchronous constraint does not allow change of $x_i$ for $T_j$, where $j \neq i$.
  The other requirements of the token ring problem are $\psi_{\mathbf{safety}}$ 
and $\psi_{\mathbf{fairness}}$, which should hold in the set of legitimate 
states.  To guarantee them, the following SMT constraints are added to the SMT 
instance:
\begin{align}
\nonumber
\forall s \in \states:\LS(s) & \; \Rightarrow \; (\exists i \in [0, n-1]: 
(\token_i(s) \; \wedge \; \forall j \neq i: \neg \token_j(s))) \\
\nonumber \forall s \in \states \; : \; \LS(s) & \; \Rightarrow \; \forall i 
\in [0, n-1] :  (\token_i(s) \wedge (s,s') \in T_i) \Rightarrow \token_{(i+1 
\Mod n)}(s')
\end{align}


\subsubsection{Synthesis of Ideal-Stabilizing Systems}
\label{subsec:ideal-syn}

We now present the constraints specific to Problem Statement 2. The only such 
constraints is related to the two \ltlr formulas $\varphi$ and $\psi$. To this 
end, we add the following to our SMT instance:
\begin{align}
\label{eq:ideal}
\forall s \in \states \; : \; \varphi_\smt \wedge \psi_\smt
\end{align}

\paragraph{\em Example} We just present $\psi_{\mathbf{LME}}$ for 
Example~3.2, as $\varphi_{\mathbf{LME}}$ is similar to 
Example~\ref{subsec:ex1}:
\begin{align} 
\nonumber
\forall s \in \states \; : \; & \Big( (\exists i \in [0, |\Pi_\topolgy|-1]: 
\token_i(s)) \; \wedge \; (\forall i \in [0, |\Pi_\topolgy|-2]: \neg 
(\token_i(s) \wedge \token_{(i+1)}(s))) \Big)
\end{align}
\begin{align} 
\nonumber
\nonumber
&  \forall s,s' \in \states \; : \; \forall i,j \in [0, |\Pi_\topolgy|-1] \; : 
\; \neg \token_i (s) \; \wedge \; (s,s') 
\in T_j \implies \lambdai(s') > \lambdai(s)  \end{align}
\begin{align}
\nonumber \forall s \in \states \; : \; \forall i \in \{0, |\Pi_\topolgy|-1] 
\; : \; \neg \token_i (s) \implies & \exists j \in [0, |\Pi_\topolgy|-1] \; : 
\\ \nonumber & \exists s' \in \states \; : \; (s,s') \in T_j
\end{align}

\noindent Note that adding a set of constraints to an SMT instance is 
equivalent to adding their conjunction.

\subsubsection{Synthesis of Monotonic-Stabilizing Systems}
\label{subsec:mon-syn}

In order to synthesize a \linebreak monotonic-stabilizing protocol, we need to 
add a constraint to guarantee that in each recovery path, each process 
executes at most once transition. In order to enforce this property, for each 
process $\proc_i$, we define a Boolean function
$$\flag_i : S \mapsto \{\tru, \fals\}$$
and include the following
constraint to the SMT instance:
\begin{align}
\label{eq:mon1} 
  \begin{array}{r@{}l}
\forall s, s' \in S \; : \; \forall i \in [0, |\Pi_\topolgy|-1] \; : 
\; & (\neg \LS(s) \wedge   (s,s') \in T_i) \implies \\ & ( 
\flag_i(s) \, \wedge \, \neg \flag_i(s'))
  \end{array}
\end{align}

\begin{align}
\label{eq:mon2}
  \begin{array}{r@{}l}
\forall s \in S \; : \; \forall i,j \in \{0 , \dots ,
|\Pi_\topolgy|-1\} \; : \; & (\neg \LS(s) \, \wedge \, i\neq j \, \wedge \, 
(s,s') \in T_j  \, \wedge  \\ &  \neg \flag_i(s)) \implies
\neg \flag_i(s')
  \end{array}
\end{align}

The above two constraints guarantee that in every path starting from a state 
in $\neg \LS$, each process executes at most once. This can easily be proved by 
contradiction. Assume that in the set of executions of the resulting protocol, 
there exists a recovery path from a state in $\neg \LS$ to a state in $\LS$, 
in which a process (assume W.L.O.G $\proc_i$) executes more than once. Based on 
Constraint~\ref{eq:mon1}, each time $\proc_i$ executes a transition, the value 
of $\flag_i$ should change from $\tru$ to $\fals$. First time, $\proc_i$ 
executes, this change in the value of $\flag_i$ happens. Also, based on 
Constraint~\ref{eq:mon2}, it is guaranteed that in the execution of any other 
process, $\flag_i$ does not change. Now, based on Constraint~\ref{eq:mon1}, in 
the second execution of $\proc_i$, $\flag_i$ should change from $\tru$ to 
$\fals$. But we concluded that $\flag_i$ is already set to $\fals$, and cannot 
be changed by the execution of any other process, which is a contradiction.


\section{Case Studies and Experimental Results}
\label{sec:results}

We used the Alloy~\cite{j12} model finder tool for our experiments. Alloy 
performs relational reasoning over quantifiers, which means that we did not 
have to unroll quantifiers over their domains. Note that Alloy may convert the 
model into a SAT instance. The results presented in this section are based on 
experiments on a machine with Intel Core i5 2.6 GHz processor with 8GB of RAM. 
We report our results in both cases of success and failure for finding a 
solution. Failure is due to the impossibility of self-, monotonic-, or 
ideal-stabilization for certain problems. All of our case studies are available 
within our tool {\sc Assess}~\cite{fb17-sss}. We present our results for self-, 
ideal-, and monotonic-stabilization in 
Sections~\ref{sec:results-ss},~\ref{sec:results-is}, and~~\ref{sec:results-ms}, 
respectively.




\subsection{Case Studies for Self-Stabilization}
\label{sec:results-ss}


\subsubsection{Self-Stabilizing Token Ring} 
\label{subsection:tokenthree}

In Example~\ref{subsec:ex1}, each process $\proc_i$ maintains a variable $x_i$ with 
domain $\{0, 1, 2\}$. The read-set of a process is its own and its neighbors' 
variables, and its write-set contains its own variable. For example, in case of 
three processes for $\proc_1$, $R_\topolgy(1) = \{x_0,x_1,x_2\}$ and 
$W_\topolgy(1)=\{x_1\}$. Token possession and mutual exclusion constraints 
follow Example~\ref{subsec:ex1}. Table~\ref{table:dijkstra3state} presents our 
results for different input settings.  We present one 
of the solutions we found for the asynchronous strong stabilizing  token ring problem in a ring of three 
processes~\footnote{\change{We manually simplified the output of Alloy for 
presentation, although this task can be also automated.}}. First, we present the 
interpretation functions for the uninterpreted local predicates.
\begin{align} \nonumber
\token_0 \Leftrightarrow x_0 = x_2, \;\;
\token_1 \Leftrightarrow x_1 \neq x_0, \;\;  
\token_2 \Leftrightarrow x_2 \neq x_1 
\end{align} 

The synthesized solution for  transition relations for each process is the 
following:
\begin{align*}
\nonumber  &\proc_0 \; : \; &(x_0 = x_2)  \;\;\; &\rightarrow \;\;\; x_0 := (x_0+1) \Mod 3& \\
&\proc_1 \; : \;&(x_1 \neq x_0)  \;\;\; &\rightarrow \;\;\;  x_1 := x_0& \\
&\proc_2 \; : \;&(x_2 \neq x_1)  \;\;\; &\rightarrow \;\;\;  x_2 := x_1&
\end{align*}
\noindent Note that our synthesized solution is identical to that of Dijkstra's 
$k$-state solution. 
We could not synthesize the three-state solution, as in this protocol, the 
token does not always circulate in one direction (it changes its circulation 
direction), but we have this constraint in $\psi_{\mathbf{fairness}}$, as 
presented in Example~\ref{subsec:ex1}.


\begin{table}[t] 
\centering 
\footnotesize
\begin{tabular}{| c | c | c | c | c|} 
\hline 
 {\bf \# of Processes} & {\bf Self-Stabilization} & {\bf Timing Model} & 
{\bf Symmetry} & {\bf Time (sec)} \\ 
[0.5ex] 
\hline \hline 
3 & strong & asynchronous & asymmetric & 4.21  \\ %
\hline 
3 & weak & asynchronous & asymmetric & 1.91  \\ %
\hline 
4 & strong & asynchronous & asymmetric & 748.81  \\ %
\hline
4 & weak & asynchronous & asymmetric & 100.03  \\ %
\hline
\end{tabular}
\caption{Results for synthesizing token ring.}
\label{table:dijkstra3state} 
\end{table}

\subsubsection{Mutual Exclusion in a Tree} 
\label{subsection:raymond}
In the second case study, the processes form a directed rooted tree, and the 
goal is to design a self-stabilizing protocol, where at each state of $\LS$, 
one and only one process is enabled. In this topology, each process $\proc_j$ 
has a variable $h_j$ with domain $\{i \; \mid \; \proc_i \text{ is a neighbor of 
} \proc_j \} \cup \{j\}$. The problem specification is 
the following:

\begin{description}

\item[Safety]  We assume each process $\proc_i$ is associated with an 
uninterpreted local predicate $\token_i$, which shows whether $\proc_i$ is 
enabled. Thus, mutual exclusion is the following formula:
$$\psi_{\mathbf{safety}} = \exists i \in [0, n-1]: (\token_i \; 
\wedge \; \forall j \neq i: \neg \token_j)$$

\item[Fairness] Each process $\proc_i$ is eventually enabled:
$$ \psi_{\mathbf{fairness}} = \forall i \in [0, n-1]: (\ltlF \token_i)$$
\end{description}
The formula, $\psi_{\mathbf{R}}$ given as input 
is $\psi_{\mathbf{R}}=\psi_{\mathbf{safety}} \; \wedge 
\; \psi_{\mathbf{fairness}}$.

\begin{table}[t] 
\centering 
\footnotesize
\begin{tabular}{| c | c | c | c |} 
\hline 
 {\bf \# of Processes} & {\bf Self-Stabilization} & {\bf Timing Model}  & {\bf Time (sec)} \\ 
[0.5ex] 
\hline \hline 
3 & strong & synchronous &  0.84  \\ %
\hline 
4 & strong & synchronous &  16.07 \\ %
\hline
4 & weak & synchronous &  26.8  \\ %
\hline 
\end{tabular} 
\caption{Results for synthesizing mutual exclusion on a tree (Raymond's 
algorithm).} 
\label{table:raymond} 

\end{table}

Using the above specification, we synthesized a synchronous self-stabilizing 
systems, which resembles Raymond's mutual exclusion algorithm on a 
tree~\cite{R89}. Table~\ref{table:raymond} shows the experimental results. 
We present one of our solutions for token circulation on a tree, where there is 
a root with two leaves. The interpretation functions for the uninterpreted local 
predicates are as follows:
\begin{align} \nonumber
\forall i  \; : \; \token_i \Leftrightarrow h_i = i 
\end{align} 

\begin{figure}[t]
 \centering
 \includegraphics[width=.9\textwidth]{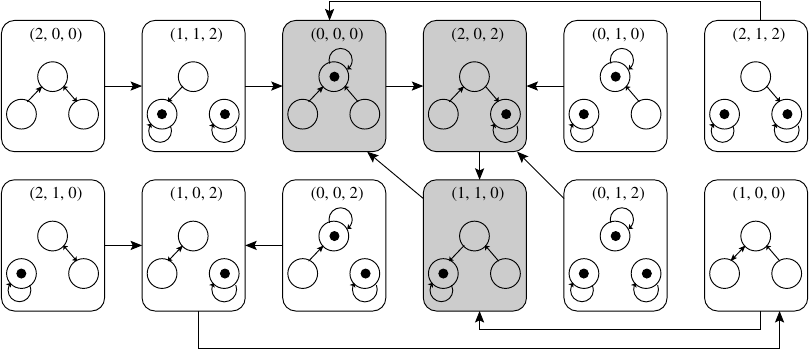}
 \caption{Self-stabilizing mutual exclusion in a tree of size 3 (Raymond's 
algorithm).}
 \label{fig:raymond}
\end{figure}

Another part of the solution is the transition relation. Assume $\proc_0$ to be 
the root process, and $\proc_1$ and $\proc_2$ to be the two leaves of the tree. 
Hence, the variable domains are $D_{h_0}=\{0,1,2\}$, $D_{h_1}=\{0,1\}$, and 
$D_{h_2}=\{0,2\}$. Fig.~\ref{fig:raymond} shows the transition relation over 
states of the form $(h_0,h_1,h_2)$ as well as pictorial representation of the 
tree and token, where the states in $\LS$ are shaded.

\subsection{Case Studies for Ideal-Stabilization}
\label{sec:results-is}

\subsubsection{Leader Election} 
\label{subsection:leader}

In leader election, a set of processes choose a leader among themselves. 
Normally, each process has a subset of states in which it is distinguished as 
the leader. In a legitimate state, exactly one process is in its leader state 
subset, whereas the states of all other processes are outside the corresponding 
subset.

We consider line and tree topologies. Each process has a variable $c_i$ and we 
consider domains of size two and three to study the existence of an 
ideal-stabilizing leader election protocol. To synthesize such a protocol, we 
associate an uninterpreted local predicate $l_i$ for each process $\proc_i$, 
whose value shows whether or not the process is the leader. Based on the 
required specification, in each state of the system, there is one and only 
one process $\proc_i$, for which $l_i=\tru$:
$$\psi_{\mathbf{safety}} = \exists i \in [0, n-1]: (l_i \; \wedge 
\; \forall j \neq i: \neg l_j)$$

\noindent The results for this case study are presented in 
Table~\ref{table:leader}. In the topology column, the structure of the 
processes along with the domain of variables is reported.  In the case of 4 
processes on a line topology and tree/2-state, no solution is found. The 
time we report in the table for these cases are the time needed to report 
unsatisfiability by Alloy.

\begin{table}[t] 
	
\centering 
\footnotesize
\begin{tabular}{| c | c | c | c|} 
\hline 
 {\bf \# of Proc.}  & {\bf Timing Model} & 
{\bf Topology} & {\bf Time (sec)} \\ 
[0.5ex] 
\hline \hline 
3 & asynchronous & line/2-state & 0.034  \\ %
\hline 
4  & asynchronous & line/2-state & 0.73  \\ %
\hline
4 & asynchronous & line/3-state & 115.21 \\
\hline 
4 & asynchronous & tree/2-state & 0.63 \\
\hline 
4 & asynchronous & tree/3-state & 12.39 \\
\hline 
\end{tabular}

\caption{Results for ideal-stabilizing leader election.} 
\label{table:leader} 
\end{table}

We present the asynchronous  solution  for the case of three processes on 
a line, where each 
process $\proc_i$ has a Boolean variable $c_i$. Since the only specification for 
this problem is state-based (safety), there is no constraint on the transition 
relations, and hence, we only present the interpretation function for each 
uninterpreted local predicate $l_i$.
\begin{align}
\nonumber
l_0  = (c_0 \wedge \neg c_1) \;\;\;\;
l_1  = (\neg c_0 \wedge \neg c_1) \, \vee \, (c_1 \wedge \neg c_2) \;\; \;\;
\label{act:leader}
l_2  = (c_1 \wedge c_2)
\end{align}
\subsubsection{Local Mutual Exclusion} 
\label{subsection:mutual}

Our next case study is local mutual exclusion, as discussed in 
Example 3.2. We consider a line topology in which each process 
$\proc_i$ has a Boolean variable $c_i$. The results for this case study are 
presented in Table~\ref{table:ideal_me}.

\begin{table}[t] 
\centering 
\footnotesize
\begin{tabular}{| c | c | c | c|} 
\hline 
 {\bf \# of Proc.} & {\bf Timing Model} & 
{\bf Symmetry} & {\bf Time (sec)} \\ 
[0.5ex] 
\hline \hline 
3  & asynchronous & asymmetric & 0.75  \\ %
\hline 
4  & asynchronous & asymmetric & 24.44  \\ %
\hline
\end{tabular} 
\caption{Results for synthesizing ideal stabilizing local mutual exclusion.} %
\label{table:ideal_me} 

\end{table}

The solution we present for  the local mutual exclusion problem corresponds to 
the case of asynchronous system with four processes on a ring. Note that for each process $\proc_i$, when 
$\token_i$ is true, the transition $T_i$ changes the value of $c_i$. Hence, 
having the interpretation functions of $\token_i$, the definition of transitions 
$T_i$ are determined as well. Below, we present the interpretation functions of 
the uninterpreted local predicates $\token_i$.
\begin{align}
\nonumber
\token_0 & = (c_0 \wedge c_1) \, \vee \, (\neg c_0 \wedge \neg c_1) \\ \nonumber
\token_1 & = (\neg c_0 \wedge c_1 \wedge c_2) \, \vee \, (c_0 \wedge \neg c_1 \wedge \neg c_2) \\ \nonumber 
\token_2 & = (\neg c_1 \wedge c_2 \wedge \neg c_3) \, \vee \, ( c_1 \wedge \neg c_2 \wedge  c_3) \\ \nonumber 
\token_3 & = (c_2 \wedge c_3) \, \vee \, (\neg c_2 \wedge \neg c_3)
\end{align}

\subsection{Case Studies for Monotonic-Stabilization}
\label{sec:results-ms}

\subsubsection{Maximal Independent Set} 
\label{subsection:mis}

Given an undirected graph $G=(V,E)$, we say that $S \subseteq V$ is an 
independent set of $G$, if no two vertices in $S$ share an edge in $E$. The 
set $S$ is a maximal independent set (MIS), if it is not a proper subset of any 
other independent set. We use a similar topology as used in the 
literature~\cite{srr95}. Assuming processes to be the vertices of the 
graph, we consider a Boolean variable $\ind_i$ for each process $\proc_i$. The 
value of $\ind_i$ determines whether or not $\proc_i$ is part of the independent 
set or not. A legitimate state is the one where processes with $\tru$ values of 
their $\ind$ variables form an independent set. For example, considering a ring 
of four processes, the set of legitimate states can be specified by the 
following predicate:

\begin{align}
\nonumber & (\ind_0(s) \wedge \neg \ind_1(s)  \wedge \ind_2(s) \wedge \neg \ind_3(s) ) 
\; \vee \\ \nonumber & (\neg \ind_0(s)  \wedge \ind_1(s)  \wedge \neg \ind_2(s)  \wedge \ind_3(s)) 
\end{align}

In this case study, our goal is to synthesize monotonic-stabilizing MIS 
protocols for ring topologies, where each process can read its own variable, as 
well as the variables of its neighbors, and can only write to its own variable. 
The results of this case study are presented in Table~\ref{table:mis}. The 
last column indicates whether or not Alloy is able to find a solution. Note 
that since our method is complete, unsatisfiability means that there exists no 
protocol satisfying the specified requirements. The following is the 
synthesized symmetric asynchronous protocol for the case of three processes in a 
ring topology. Note that in the case of symmetric protocol, all processes 
execute similarly.

\begin{table}[t] 
	
\centering 
\footnotesize
\begin{tabular}{| c | c | c |  c| c|} 
\hline 
 {\bf \# of Proc.}  & {\bf Timing Model} & {\bf Symmetry}  & {\bf Time (sec)}  & {\bf Result}\\ 
[0.5ex] 
\hline \hline 
3 & asynchronous & asymmetric  & 0.35 & sat \\ %
\hline 
3 & asynchronous & symmetric  & 0.09 & sat \\ %
\hline 
3 & synchronous & asymmetric  & 0.08 & sat \\ %
\hline 
4 & asynchronous & asymmetric  & 1.35  & sat \\ %
\hline 
4 & asynchronous & symmetric  & 0.9  & unsat \\ %
\hline 
5 & asynchronous & asymmetric & 6.33  & sat \\ %
\hline 
6 & asynchronous & asymmetric  & 63.24 & sat  \\ %
\hline
\end{tabular}

\caption{Results for monotonic stabilizing maximal independent set in ring.} 
\label{table:mis} 
\end{table}

 \begin{align*}
 \nonumber  
 &\proc_i \; : \; &\neg \ind_i \wedge \neg \ind_{l} \wedge \neg \ind_{r} \;\;\; &\rightarrow \;\;\; \ind_i := \tru& \\
 && \ind_i \wedge  \ind_{l} \;\;\; &\rightarrow \;\;\; \ind_i := \fals&
 \end{align*}

\noindent \noindent In the above synthesized protocol, $r$ is the index of the 
right process, or  $r=(i+1) \, \mod \, 3$, and $l$ is the index of the left 
process, or  $l=(i-1) \, \mod \, 3$. With a simple observation of the above 
synthesized protocol, we can see that in any path starting from a non-legitimate 
state, each process takes at most one action. We also present one of the 
solutions for the case of 4 processes in a ring topology.

\begin{align*}
\nonumber  &\proc_0 \; : \; &\ind_0 \wedge \ind_3 \;\;\; &\rightarrow \;\;\; \ind_0 := \fals& \\
&\proc_1 \; : \; &\neg \ind_1 \wedge \neg \ind_2\;\;\; &\rightarrow \;\;\; \ind_1 := \tru& \\
&\proc_2 \; : \; &\ind_2 \wedge \ind_3 \;\;\; &\rightarrow \;\;\; \ind_2 := \fals & \\
&& \ind_2 \wedge \neg \ind_3 \wedge \ind_1 \;\;\; &\rightarrow \;\;\; \ind_2 := \fals & \\
&\proc_3 \; : \; &\neg \ind_3 \wedge \neg \ind_0 \;\;\; &\rightarrow \;\;\; \ind_3 := \tru & \\
&&\neg \ind_3 \wedge \ind_0 \wedge \neg \ind_2 \;\;\; &\rightarrow \;\;\; \ind_3 := \tru &
\end{align*}

\subsubsection{Maximal Independent Set in Unidirectional Rings} 
\label{subsection:mis2}

Yamauchi and Tixeuil~\cite{YT10} state that monotonic stabilization requires 
additional information exchange between processes. In our second case study, we 
attempt to limit information exchange in maximal independent set and see 
whether we can still synthesize monotonic-stabilizing protocols for this 
problem. We considered unidirectional rings for this case study. In other words, 
each process can only read its own variable and the variable of its left 
process, and can write to its own variable. For example, for a ring of three 
processes, $R_{\proc_0}=\{ \ind_0 , \ind_2 \}$ and $W_{\proc_0}=\{ \ind_0 \}$. 
The results for this case study are presented in Table~\ref{table:mis2}. As can 
be seen, for the case of asymmetric asynchronous topologies, a protocol 
is found for rings of even size (4 and 6), but not for rings of odd size (3 and 
5). Although, our solution is not general, but it can give an intuition 
to protocol designers for a general monotonic-stabilizing protocol to solve 
maximal independent in unidirectional rings. For the case of synchronous 
protocol with three processes, we synthesized the following solution:

\begin{align*}
\nonumber  &\proc_0 \; : \; &\tru \;\;\; &\rightarrow \;\;\; \ind_0 := \fals& \\
&\proc_1 \; : \; &\tru \;\;\; &\rightarrow \;\;\; \ind_1 := \fals& \\
&\proc_2 \; : \; &\tru \;\;\; &\rightarrow \;\;\; \ind_2 := \tru& \\
\end{align*}

As can be simply observed, the synthesized topology takes every state 
(legitimate or non-legitimate) to one state $\left< \fals,\fals,\tru \right>$. 
A question that may raise for the reader is that why a similar protocol does not 
work in the case of asynchronous systems. The answer is that in the case of 
asynchronous systems, each step of the system is the execution of exactly one 
process, and hence, one execution of such a protocol may take the system from 
$\LS$ to a non-legitimate state (closure violation). For example, asynchronous 
execution of the synthesized actions for the synchronous case will take the 
system from $\left<\tru,\fals,\fals\right>$, which is legitimate to 
$\left<\fals,\fals,\fals\right>$, which is non-legitimate (the first action is 
taken).

\begin{table}[t] 
	
\centering 
\footnotesize
\begin{tabular}{| c | c | c |  c| c|} 
\hline 
 {\bf \# of Proc.}  & {\bf Timing Model} & {\bf Symmetry}  & {\bf Time (sec)}  & {\bf Result}\\ 
[0.5ex] 
\hline \hline 
3 & asynchronous & asymmetric  & 0.06 & unsat \\ %
\hline 
3 & asynchronous & symmetric  & 0.05 & unsat \\ %
\hline 
3 & synchronous & asymmetric  & 0.07 & sat \\ %
\hline 
4 & asynchronous & asymmetric  & 0.74  & sat \\ %
\hline 
4 & asynchronous & symmetric  & 0.44  & unsat \\ %
\hline 
5 & asynchronous & asymmetric & 8.73  & unsat \\ %
\hline 
6 & asynchronous & asymmetric  & 85.14 & sat \\ %
\hline
\end{tabular}

\caption{Results for monotonic stabilizing maximal independent set in unidirectional ring.} 
\label{table:mis2} 
\end{table}

\subsubsection{Grundy Coloring}

Our third case study is the problem of Grundy coloring. Considering a graph 
$G=(V,E)$, and a coloring function $$\col : V \rightarrow [1, k]$$ a vertex $v 
\in V$ is called a {\em Grundy} node if
$$\col(v) = \min \big\{l \in [1, k] \mid \forall u: (v,u) \in E \implies ( 
\col(u) \neq l)\big\}$$
Simply speaking, $v$ is colored with the smallest color not taken
by any neighbor.  A Grundy coloring for a graph is one in which every node
is a Grundy node. 

To synthesize a monotonic-stabilizing protocol for this problem, we consider a 
set of processes as the nodes of the graph, such that each process has a $\col$ 
variable. The designer can specify the domain of the $\col$ variables. Each 
process can read its own variable, and the variables of its neighbors, and can 
write to its own variable. The set of legitimate states are those, in which each 
process is a Grundy node. For example, considering a ring of 4 processes, where 
the domain of $\col$ variables is $\{1,2,3\}$, the set of legitimate states can 
be specified by the following predicate:

\begin{align}
\nonumber  \forall i \in \{0,1,2,3 \} : \; & \big(\col_i(s) \neq \col_{(i+1  
\!\mod 4)}(s)\big) \; \wedge \\
\nonumber & \big(\col_i(s)=2 \implies (\col_{(i+1 \!\mod 4)}(s)=1 \, \vee \, 
\col_{(i-1 \!\mod 4)}(s)=1)\big) \; \wedge \\
\nonumber & \big(\col_i(s)=3 \implies ((\col_{(i+1 \!\mod 4)}(s)=2 \, \vee \,
\col_{(i-1 \!\mod 4)}(s)=2) \; \wedge \; \\
\nonumber & \hspace{3.5cm} (\col_{(i+1 \mod 4)}(s)=1 \, \vee \, \col_{(i-1 \!\mod 
4)}(s)=1) )\big)
\end{align}

\noindent  Note that the last three lines of the above predicate are ensuring 
that the assigned color to each node is the minimum available one. Our results 
for this case study are presented in Table~\ref{table:grundy}. Our synthesized 
protocol for the case of a symmetric protocol with three processes in a ring is 
the following:
\begin{align*}
\nonumber  
&\proc_i \; : \; & (\col_i = 1) \wedge (\col_{l} \neq 2) \wedge (\col_{r} =1) \;\;\; &\rightarrow \;\;\; \col_i := 2& \\
&& (\col_i = 1) \wedge (\col_{l} =2) \wedge (\col_{r} =1) \;\;\; &\rightarrow \;\;\; \col_i := 3& \\
&&(\col_i = 3) \wedge (\col_{l} =3) \wedge (\col_{r} \neq 2) \;\;\; &\rightarrow \;\;\; \col_i := 2 & \\
&&(\col_i = 3) \wedge (\col_{l} =2) \wedge (\col_{r} =3) \;\;\; &\rightarrow \;\;\; \col_i := 1 & \\
&&(\col_i = 2) \wedge (\col_{l} =2) \wedge (\col_{r} =3) \;\;\; &\rightarrow \;\;\; \col_i := 1 & \\
&&(\col_i = 2) \wedge (\col_{l} \neq 3) \wedge (\col_{r} = 2) \;\;\; &\rightarrow \;\;\; \col_i := 3 & 
\end{align*}
 
\noindent In the above synthesized protocol, $r$ is the index of the right 
process, or  $r=(i+1) \, \mod \, 3$, and $l$ is the index of the left process, 
or  $l=(i-1) \, \mod \, 3$. Note that in this case, Grundy coloring is the same 
as the three-coloring problem~\cite{ga09}.

We also present our synthesized model for the case of asynchronous protocol with 
4 processes in a line topology:
 
 \begin{align*}
 \nonumber  
 &\proc_0 \; : \; & (\col_0 \neq 1) \wedge (\col_{1} = 3)  \;\;\; &\rightarrow \;\;\; \col_0 := 1& \\
 && (\col_0 = 3) \wedge (\col_{1} = 1)  \;\;\; &\rightarrow \;\;\; \col_0 := 2& \\
&\proc_1 \; : \; & (\col_0 = 3) \wedge (\col_{1} = 2)  \;\;\; &\rightarrow \;\;\; \col_1 := 3& \\
& & (\col_0 = 2) \wedge (\col_{1} = 2)  \;\;\; &\rightarrow \;\;\; \col_1 := 1& \\
& & (\col_0 = 1) \wedge (\col_{1} \neq 3) \wedge (\col_{2} = 2)  \;\;\; &\rightarrow \;\;\; \col_1 := 3& \\
& & (\col_0 = 1) \wedge (\col_{1} =1) \wedge (\col_{2} \neq 2)  \;\;\; &\rightarrow \;\;\; \col_1 := 3& \\
&\proc_2 \; : \; & (\col_1 = 3) \wedge (\col_{2} \neq 2) \wedge (\col_3=1)  \;\;\; &\rightarrow \;\;\; \col_2 := 2& \\
&& (\col_1 \neq 2) \wedge (\col_{2} = 1) \wedge (\col_{3} = 2) \;\;\; &\rightarrow \;\;\; \col_2 := 2& \\
&& (\col_1 \neq 3) \wedge (\col_{2} =1) \wedge (\col_{3} = 1) \;\;\; &\rightarrow \;\;\; \col_2 := 2& \\
&& (\col_1 = 1) \wedge (\col_{2} =3) \wedge (\col_{3} \neq 2) \;\;\; &\rightarrow \;\;\; \col_2 := 2& \\
&\proc_3 \; : \; & (\col_2 \neq 1) \wedge (\col_{3} \neq 1)  \;\;\; &\rightarrow \;\;\; \col_3 := 1& \\
&& (\col_2 = 1) \wedge (\col_{3} =3)  \;\;\; &\rightarrow \;\;\; \col_3 := 1& 
 \end{align*}

\begin{table}[t] 
	
	\centering 
\footnotesize
\begin{tabular}{| c | c | c |  c| c| c |} 
\hline 
 {\bf \# of Proc.}  & {\bf Timing Model} & {\bf Topology} & {\bf Symmetry}  & {\bf Time (sec)}  & {\bf Result}\\ 
[0.5ex] 
\hline \hline 
3 & asynchronous & ring & asymmetric  & 3.02 &  sat \\ %
\hline 
3 & asynchronous & ring & symmetric  & 2.89 & sat \\ %
\hline 
3 & synchronous & ring & asymmetric  & 4.29 & sat \\ %
\hline 
3 & asynchronous & line & asymmetric  & 3.93 & sat  \\ %
\hline 
4 & asynchronous & ring & asymmetric  & 102.46  & sat \\ %
\hline 
4 & asynchronous & ring & symmetric  & 152.69  & unsat \\ %
\hline
4 & asynchronous & line & asymmetric  & 144.16  & sat \\ %
\hline
\end{tabular}

\caption{Results for monotonic stabilizing Grundy coloring.} 
\label{table:grundy} 
\end{table}


\section{Discussion}
\label{sec:disc}

In this section, we address a few points often raised about this line of work.

\subsection{Applicability}

First, notice that a user of our technique has to 
give the network topology in terms of read/write restrictions as input. This in 
turn means that the user has to choose a set of variables and their domains. 
Although choosing variables and their domains may seem challenging, in most 
cases, the user can have educated guesses about the variables and their domains 
from the specification (similar to programming practices), as in the maximal 
matching example. This process may involve some trial and error though. For 
example,  one can start by assigning each process a Boolean variable, and if no 
solution is found, increase the number of variables or their domains (increase 
the local state space of each process).

\subsection{Scalability} 

It is obvious that scalability is an issue in this 
method. However, note that although our case studies deal with synthesizing a 
small number of processes (due to the high complexity of synthesis), having 
access to a solution for a small number of processes may give key insights to 
designers of self-stabilizing protocols to generalize the protocol for any 
number of processes. For example, our method can be applied in cases where 
there exists a cut-off point~\cite{jb14}, and we can theoretically prove that 
the solution works for any number of processes. Also, in cases, where we find 
that there is no solution for the problem, this may be a hint for a general 
impossibility result. 

One way to improve scalability is by using a counterexample-guided inductive 
synthesis loop, where an over-approximation is quickly synthesized and then 
later refined by identifying counterexamples. 

\subsection{The Choice of SMT-solver}

It is noteworthy to mention that we have conducted experiments using 
Z3~\cite{mb08} and Yices~\cite{d14} SMT solvers as well, and in the majority of 
our cases studies, Alloy was the fastest model solver. We should also mention 
that the maximum number of processes in the system we could synthesize differs 
from problem to problem. This number solely depends on the complexity of the 
input specification and, hence, the SMT instance. That means there is no fixed 
maximum number of processes that this method can handle. Note that the maximum 
number reported in this paper is the maximum number of processes we could find 
a solution for each case study in less than an hour. 

\subsection{Synthesis under Worst-Case Recovery Time Constraint}
\label{subsec:recovery}
 
There are quantitative metrics in stabilizing systems that are as crucial as 
closure and convergence in practice. One of these metrics is {\em recovery 
time}, which is essentially the length of the path starting in a state in $\neg 
\LS$ and ending in a state in $\LS$. Recovery time is crucial in designing 
stabilizing systems for some applications, such as in network protocols. In 
such applications, it may not be desirable for the recovery time to exceed a 
specific number of steps, say $w$. Thus, we can include a constraint based on 
the ``worst-case recovery time'' to our model. The constraint is the following:
\begin{equation}
\label{eq:rec}
\forall s \in \states: (0 \le \mylambda(s)) \wedge (\mylambda(s) \le w) 
\end{equation}

Based on the above constraint, since $\mylambda$ is incremented in every 
step, and its range is $w$, the worst-case recovery time in the synthesized 
system cannot exceed $w$. 


\section{Related Work}
\label{sec:related}

\subsection{Bounded Synthesis}

In bounded synthesis~\cite{fs13}, given is a set 
of \ltl\ properties, a system architecture, and a set of bounds on the size of 
process implementations and their composition. The goal is to synthesize an 
implementation for each process, such that their composition satisfies the given 
specification. The properties are translated to a universal co-B\"{u}chi 
automaton, and then a set of SMT constraints are derived from the automaton. 
Our work is inspired by this idea for finding the SMT constraints for 
strong convergence and also the specification of legitimate states. For other 
constraints, such as the ones for synthesis of weak convergence, asynchronous 
and symmetric systems, we used a different approach from bounded synthesis. The 
other difference is that the main idea in bounded synthesis is to put a bound on 
the number of states in the resulting state-transition systems, and then 
increase the bound if a solution is not found. In our work, since the purpose is 
to synthesize a self-stabilizing system, the bound is the number of all possible 
states, derived from the given topology. 
 
\subsection{Synthesis of Self-Stabilizing Systems}

In~\cite{ke13}, the authors show that adding strong convergence is 
NP-complete in the size of the state space, which itself is exponential in the 
number of variables of the protocol. Ebnenasir and Farahat~\cite{ef11} 
also proposed an automated method to synthesize self-stabilizing algorithms. Our 
work is different in that the method in~\cite{ef11} is not complete for strong 
self-stabilization. This means that if it cannot find a solution, it does not 
necessarily imply that there does not exist one. However, in our method, if the 
SMT-solver declares ``unsatisfiability'', it means that no self-stabilizing 
algorithm that satisfies the given input constraints exists.\removeCR{ Also, 
using our approach, one can synthesize synchronous and asynchronous programs, 
while the method in~\cite{ef11} synthesizes asynchronous systems only. Finally, 
our method is based on the constantly-evolving technique of SMT solving. We 
expect our technique to become more efficient as more efficient SMT solvers 
emerge.} A complete synthesis technique for self-stabilizing systems is 
introduced in~\cite{ke14}. The limitations of this work compared to ours is: (1) 
unlike the approach in~\cite{ke14}, we do not need the explicit description of 
the set of legitimate states, and (2) the method in~\cite{ke14} needs the set of 
actions on the underlying variables in the legitimate states. We also emphasize 
that although our experimental results deal with small numbers of processes, 
our approach can give key insights to designers of self-stabilizing protocols 
to generalize the protocol for any number of processes~\cite{jb14}.

Another line of research is the work in~\cite{bbj16}. The authors in 
this paper also introduce a technique to synthesize self-stabilizing protocols 
based on bounded synthesis, but their main focus is on Byzantine failures. To 
this end, they use a counterexample-guided inductive synthesis loop for 
networks of fixed size.

\subsection{Automated Addition of Fault-Tolerance}

The proposed algorithm in~\cite{bka12} synthesizes a fault-tolerant distributed 
algorithm from its fault-intolerant version. The distinction of our work with 
this study is (1) we emphasize on self-stabilizing systems, where any system 
state could be reachable due to the occurrence of any possible fault, (2) the 
input to our problem is just a system topology, and not a fault-intolerant 
system, and (3), the proposed algorithm in~\cite{bka12} is not complete. 
Bonakdarpour and Kulkarni studied the complexity of synthesizing timed 
multi-phased fault recovery in~\cite{bk15}. Finally, we introduced efficient 
symbolic heuristics for timed multi-phase recovery in~\cite{fb17}.
  

\section{Conclusion}
\label{sec:conclusion}

In this paper, we proposed an automated SMT-based technique for synthesizing 
self-, ideal-, and monotonic-stabilizing algorithms. The required input to our 
approach is a high-level specification of the algorithm, given in the linear 
temporal logic (LTL) and the network topology. In the particular case of 
self-stabilization, this means that the detailed description of the set of 
legitimate states is not required. This relaxation is significantly beneficial, 
as developing a detailed predicate for legitimate states can be a tedious task. 
Our approach is sound and complete for finite-state systems; i.e., it ensures 
correctness by construction and if it cannot find a solution, we are guaranteed 
that there does not exist one. We demonstrated the effectiveness of our 
approach by automatically synthesizing Dijkstra's token ring, Raymond's mutual 
exclusion, and ideal-stabilizing leader election and local mutual exclusion 
algorithms as well as monotonic-stabilizing maximal independent set and Grundy 
coloring.

We note that our approach can be easily extended to incorporate additional 
properties of self-stabilizing systems. For instance, one can impose a 
worst-case recovery time constraint by putting an upperbound on the number of 
recovery steps. This can be simply achieved by including a constraint on the 
$\mylambda$ function (i.e., Constraint~\ref{eq:rec}).   

For future, we plan to work on synthesis of probabilistic self-stabilizing 
systems. Another challenging research direction is to devise synthesis 
methods where the number of distributed processes is parameterized as well as 
cases where the size of state space of processes is infinite. We note that 
parameterized synthesis of distributed systems, when there is a cut-off point 
is studied in~\cite{jb14}. Our goal is to study parameterized synthesis for 
self-stabilizing systems, and we plan to propose a general method that works not 
just for cases with cut-off points. We would also like to investigate the 
application of techniques such as counterexample-guided inductive synthesis to 
improve the scalability of the synthesis process.

\section{Acknowledgments}

This research was supported in part by Canada NSERC Discovery Grant 418396-2012
and NSERC Strategic Grant 430575-2012. We would also like to acknowledge the 
anonymous referees who carefully reviewed the article and provided us with 
many constructive comments.  

\bibliographystyle{plain}
\bibliography{main}
\end{document}